\documentclass[aps,twocolumn,superscriptaddress,showpacs]{revtex4}
\usepackage{epsfig}
\usepackage{color}
\usepackage{graphicx}
\usepackage{dcolumn}
\usepackage{bm}
\usepackage[normalem]{ulem}

\begin{document}

\title{Correlation strength, Lifshitz transition and the emergence of a two- to three-dimensional crossover in FeSe under pressure}

\author{S. L. Skornyakov}

\affiliation{Institute of Metal Physics, Sofia Kovalevskaya Street 18, 620219 Yekaterinburg GSP-170, Russia}
\affiliation{Ural Federal University, 620002 Yekaterinburg, Russia}

\author{V. I. Anisimov}

\affiliation{Institute of Metal Physics, Sofia Kovalevskaya Street 18, 620219 Yekaterinburg GSP-170, Russia}
\affiliation{Ural Federal University, 620002 Yekaterinburg, Russia}

\author{D. Vollhardt}

\affiliation{Theoretical Physics III, Center for Electronic Correlations and
Magnetism, Institute of Physics, University of Augsburg, 86135 Augsburg, Germany}

\author{I. Leonov}

\affiliation{Institute of Metal Physics, Sofia Kovalevskaya Street 18, 620219 Yekaterinburg GSP-170, Russia}
\affiliation{Materials Modeling and Development Laboratory, National University of Science and Technology 'MISIS', 119049 Moscow, Russia}

\date{\today}

\begin{abstract}
We report a detailed theoretical study of the electronic
structure, spectral properties, and lattice parameters of
bulk FeSe under pressure using a fully charge
self-consistent implementation of the density functional
theory plus dynamical mean-field theory method (DFT+DMFT).
In particular, we perform a structural optimization and
compute the evolution of the lattice parameters (volume,
$c/a$ ratio, and the internal $z$ position of Se) and the
electronic structure of the tetragonal (space group $P4/nmm$)
unit cell of paramagnetic FeSe. Our results for the lattice
parameters obtained by structural optimization of FeSe
using DFT+DMFT are in good quantitative agreement with
experiment. The $c/a$ ratio is slightly overestimated 
by about $3$~\%, presumably due to the absence of the
van der Waals interactions between the FeSe layers in our
calculations. The lattice parameters determined
within nonmagnetic DFT are off the experimental values
by a remarkable $\sim$$6$-$15$~\%, implying a crucial importance
of electron correlations in determining the
correct lattice properties of FeSe. Upon compression to
$10$~GPa, the $c/a$ ratio and the lattice volume show a
linear-like decrease by $2$ and $10$~\%, respectively,
while the Se $z$ coordinate weakly increases by $\sim$$2$~\%.
Most importantly, our results reveal a topological change
of the Fermi surface (Lifshitz transition) which is
accompanied by a two- to three-dimensional crossover. Our
results indicate a small reduction of the quasiparticle
mass renormalization $m^*/m$ by about $5$~\% for the $e$
and less than $1$~\% for the $t_2$ states, as compared to
ambient pressure. The behavior of the momentum-resolved
magnetic susceptibility $\chi({\bf q})$ shows no topological
changes of magnetic correlations under pressure, but
demonstrates a reduction of the degree of the in-plane
$(\pi,\pi)$ stripe-type nesting. Our results for the
electronic structure and lattice parameters of FeSe are in
good qualitative agreement with recent experiments on its
isoelectronic counterpart FeSe$_{1-x}$S$_x$.
\end{abstract}

 \maketitle


\section{Introduction}

Iron-chalcogenides (FeCh) with the PbO crystal structure
have attracted much attention recently due to their anomalous
properties caused by the complex interplay between electronic,
magnetic, and lattice degrees of freedom~\cite{review_chalcogen_recent}.
In this context, FeSe is one of the most well-known
members of the so-called 11-family, which shows
remarkable connections between magnetism,
nematicity, and superconductivity. FeSe has the same
planar structure of the FeCh tetrahedra planes as
the iron pnictides, bound together by the van der
Waals interaction, but without separating
layers~\cite{pnictide_discovery,review_fe_supercond_general}.
In fact, it is structurally the simplest among other
Fe-based superconductors, with a superconducting phase
emerging below a critical temperature $T_c$$\sim$$8$~K
without applying doping and/or external
pressure~\cite{Superconductivity_FeSe}. In contrast
to iron pnictides, FeSe shows no long-range magnetic
order, but displays a strong enhancement of short-range
spin fluctuations near $T_c$, with a resonance
at the $(\pi,\pi)$ magnetic vector (stripe-type) in
the spin excitation spectra~\cite{FeSe_magn_sc_pressure}.
Moreover, similar to many other Fe-based materials, it
exhibits a tetragonal-to-orthorhombic
transition to a nematic phase below $T_s$$\sim$$90$~K,
with a spontaneously broken lattice rotational $C_4$
symmetry in the $ab$ plane~\cite{review_chalcogen_recent,FeSe_structure,FeSe_tetr_ort}.

Upon compression, $T_s$ appears to decrease, followed
by the emergence of a magnetically long-range ordered
phase at $\sim$$1$~GPa~\cite{FeSe_magn_1GPa}. It leads
to a dramatic increase of the critical temperature $T_c$
to a maximum of about $37$~K at $\sim$$6$~GPa, implying
that $T_c$ depends very sensitively on even a moderate
change of the lattice volume~\cite{FeSe_magn_sc_pressure,FeSe_sc_enhance_press}.
Furthermore, it has been recently shown that the electronic
and lattice properties of bulk FeSe can be effectively
tuned by the isoelectronic substitution of Se by S and Te, respectively. Namely, 
S has a smaller ionic radius than Se and hence results
in a compression of the unit cell, while Te has a larger ionic radius,
leading to an expansion of the lattice. In the expanded lattice  $T_c$ is found
to increase up to $14$~K~\cite{Tc_negative_pressure},
while tetragonal FeS has been reported to be a superconductor
with $T_c$$\sim$$5$~K~\cite{FeS_superconductivity}. Moreover,
superconductivity in FeSe can be boosted to $\sim$$40$~K
and even to about $100$~K by means of K-intercalation~\cite{FeSe_intercalation}
and in the case of a monolayer of FeSe on SrTiO$_3$~\cite{FeSe_monolayer}.
Interestingly, while FeSe is not magnetically long-range ordered, magnetism appears both 
upon compression (most probably consistent
with a stripe-type $(\pi, \pi)$ vector~\cite{FeSe_pi_pi})
and upon expansion. For example,
the end-member FeTe has a long-range $(\pi,0)$ antiferromagnetic
order~\cite{FeTe_pi_0}. These findings clearly
demonstrate that isoelectronic tuning of the lattice of
FeSe, e.g., by doping, can be an effective means to
control $T_c$ in FeCh.

On the experimental and theoretical sides, much effort has 
been devoted to understand the electronic properties of FeCh 
employing, e.g., angle-resolved photoemission (ARPES) 
measurements~\cite{review_chalcogen_recent} and theoretical 
band structure calculations~\cite{FeSe_dft}. In particular, 
it has been shown that FeSe has the same Fermi surface (FS) 
topology as the pnictides, characterized by an in-plane 
stripe-type nesting $(\pi, \pi)$, consistent with $s^\pm$ 
pairing symmetry~\cite{FeSe_spm}. A significant narrowing of 
the Fe~$3d$ bandwidth by a factor of $\sim$$2$ and a large 
orbital-selective enhancement of the quasiparticle mass 
$m^*/m$ in the range of $\sim$$3-20$ was 
reported~\cite{FeSe_arpes_1,FeSeTe_arpes}, implying a crucial 
importance of (orbital-selective) electronic 
correlations~\cite{FeSe_renorm_1,FeSe_LHB_1,FeSe_LHB_2,FeSe_ours,FeSe_arpes_2}. 
In addition, recent photoemission data show a formation of 
the lower Hubbard-band satellite at about $-2$~eV in bulk 
FeSe which is absent in density functional theory 
(DFT)~\cite{FeSe_arpes_2,FeSe_LHB_2}. Furthermore, computations 
of the lattice parameters of iron-based superconductors within 
DFT highlight peculiar deviations with experiments, not 
captured by standard band structure 
methods~\cite{FeSe_V_underestimation,FeSe_ours}. All this 
suggests a complex interplay between electronic correlations 
and the lattice, implying the crucial importance of the effect 
of electronic correlations in FeCh.

State-of-the-art methods for the calculation of the
electronic structure of correlated materials using band
structure methods combined with dynamical mean-field
theory (DFT+DMFT)~\cite{dmft,dftdmft_nsc} have shown to
provide a good quantitative description of the electronic
structure of FeCh~\cite{FeSe_ours,FeSe_Haule_1,FeS_dft_dmft,FeSe_LHB_2}.
In fact, recent DFT+DMFT calculations show a significant
orbital-dependent mass enhancement in the range of $2-5$
and demonstrate the existence of a lower Hubbard band at
about $-1.5$ to $-2.0$~eV below the Fermi level of
FeSe~\cite{FeSe_ours,FeSe_Haule_1,FeSe_arpes_2,FeSe_LHB_2}.
Electronic correlations in FeSe show a strong sensitivity
to variation of the crystal structure parameters. For
example, both $T_c$ and the correlation strength
are found to be affected by an isostructural change of
the lattice volume~\cite{FeSeTe_correlations}. Indeed,
upon expansion of the lattice caused by substitution of
Se for Te, $T_c$ increases up to $\sim$$14$~K. It also leads 
to a pronounced damping of quasiparticle
coherence, i.e., to an enhancement of the strength of
electronic correlations~\cite{FeSeTe_correlations}.
By contrast, upon compression of the lattice, e.g., by
doping of FeSe with S, the electronic correlations
becomes weaker, as can be expected for a correlated 
metal~\cite{FeS_corr_suppress}. This is associated with 
a substantial increase of the band width, i.e., 
tetragonal FeS is less correlated than FeSe. We also note 
that $T_c$ is about $5$~K in bulk FeS, which is close
to $\sim$$8$~K in FeSe~\cite{FeS_superconductivity}.
Interestingly, it has been recently proposed that in FeS 
the electron-phonon interaction is significant~\cite{FeS_dft_phonon}.

Moreover, the above-mentioned changes of the lattice
volume have a strong effect on the FS. In fact, in
Fe(Se$_{1-x}$Te$_x$), i.e., upon expansion of the lattice,
experiments observe a suppression of the spectral weight 
associated with the FS pockets near the M point of the 
Brillouin zone \cite{FeSeTe_arpes}. This behavior is 
accompanied by the appearance of spectral weight near 
the X point, indicating the existence of a topological 
change of the FS (Lifshitz transition). We note that 
this effect as well as a $(\pi,\pi)$ to $(\pi,0)$ 
reconstruction of magnetic correlations upon going from 
FeSe to FeTe, has been well established by recent DFT+DMFT 
calculations~\cite{FeSe_ours}. Furthermore, compression 
of the lattice due to chemical (positive) pressure cause 
a damping of the spectral weight associated with the 
inner-most FS hole pocket at the $\Gamma$ point~\cite{Watson_prb_15}. 
This suggests the possibility of superconductivity in 
FeCh without coherent quasiparticles in a hole band forming 
the inner cylinder of the FS. Overall, these results reveal 
a subtle interplay between the electronic, magnetic and 
structural properties of FeSe and demonstrate the crucial 
importance of electronic correlations.

In our previous reports~\cite{FeSe_ours} based on the 
DFT+DMFT approach, we studied the electronic structure 
and phase stability of FeSe under \emph{negative} pressure 
which can be achieved experimentally by doping FeSe 
with Te. In agreement with experiment FeSe was shown to 
exhibit a remarkable change of the electronic structure 
and the magnetic and lattice properties upon expansion 
of the lattice. Namely, a structural transition from a 
collapsed-tetragonal to tetragonal phase was found to 
occur, which is accompanied by a reconstruction of the 
electronic structure (Lifshitz transition) and by a
$(\pi,\pi)$ to $(\pi,0)$ change of magnetic correlations. 
Furthermore there is a crossover from Fermi-liquid to 
non-Fermi-liquid behavior~\cite{FeSe_ours}. At the same time
the synergy of electronic structure, magnetism, and the 
lattice of FeSe under \emph{external} pressure has remained
essentially unexplored. For that reason we will address 
this question in our present investigation to provide 
a microscopic explanation of the properties of FeSe 
under pressure.

In this paper, we focus on the interplay between the 
electronic structure and the magnetic and structural 
properties of the tetragonal ($P4/nmm$) FeSe under 
external (positive) pressure. In particular, we study 
the effect of pressure on the unit cell shape, band 
structure, magnetic properties, and the Fermi surface 
of FeSe, employing a fully charge self-consistent 
implementation of the DFT+DMFT method~\cite{dftdmft_sc}.
We perform a structural optimization and compute the 
evolution of the lattice parameters, i.e., of the volume, 
the $c/a$ ratio, the internal $z$ position of Se (which 
is related to the height of Se above the Fe plane), and 
the electronic structure of the tetragonal unit cell 
(space group $P4/nmm$) of paramagnetic FeSe. Our results 
demonstrate that electronic correlations need to be taken 
into account in the structural optimization of tetragonal 
FeSe at ambient and positive pressure. We show that the 
lattice parameters of FeSe obtained within DFT+DMFT,
i.e., when taking into account the effect of electronic
correlations, are in good agreement with experimental 
data. Most importantly, our results reveal a topological 
change of the Fermi surface (Lifshitz transition) which 
is accompanied by a two- to three-dimensional crossover. 
We obtain a small reduction of the quasiparticle mass 
renormalization $m^*/m$ by about $5$~\% for the $e$ and 
less than $1$~\% for the $t_2$ states, as compared to 
ambient pressure. While the behavior of the momentum-resolved 
static susceptibility $\chi({\bf q})$ shows no topological 
changes of magnetic correlations under pressure, it 
demonstrates a reduction of the degree of the in-plane 
$(\pi,\pi)$ stripe-type nesting. Our results for the 
electronic structure and lattice parameters of FeSe under 
pressure are in line with recent experiments on its 
isoelectronic and isostructural counterpart FeSe$_{1-x}$S$_x$. 
Overall, our results show the sensitivity of the electronic 
structure and magnetic properties of FeCh with respect to 
a small variation of the lattice. This is associated with 
a peculiarity of its correlated band structure, namely 
the proximity of a van Hove singularity associated with 
the $xy$ and $xz/yz$ Fe $3d$ orbitals to the Fermi 
level~\cite{FeSe_ours, FeSe_arpes_recent}.


\section{Method}

To study
the electronic structure, spectral properties, and phase
stability of bulk FeSe under pressure we make use of the 
DFT+DMFT computational approach~\cite{dftdmft_nsc}. We
employ a fully charge self-consistent DFT+DMFT scheme~\cite{dftdmft_sc}
and perform direct structural optimization of the tetragonal
(space group $P4/nmm$) unit cell of FeSe, calculating the
total energy as a function of volume, lattice parameter
$c/a$, and the internal $z_{\mathrm{Se}}$ position.
For simplicity $z_{\mathrm{Se}}$ was fixed to its experimental 
value $z_{\mathrm{Se}}=0.266$~\cite{FeSe_structure} in some 
of these calculations. In order to determine the pressure, 
we fit our total-energy results to the third-order Birch-Murnaghan 
equation of state~\cite{Birch}. The calculations are performed 
for the tetragonal ($P4/nmm$) crystal structure of paramagnetic 
FeSe at an electronic temperature $T = 290$~K. We note that 
according to experimental data, bulk FeSe undergoes a 
tetragonal-to-orthorhombic (nematic) phase transition at 
sufficiently lower temperature $\sim$$90$~K (at ambient 
pressure)~\cite{review_chalcogen_recent,FeSe_structure,FeSe_tetr_ort}.

The DFT+DMFT approach is implemented within plane-wave
pseudopotentials with the generalized gradient approximation
in DFT~\cite{GGA}. For the partially filled Fe $3d$ and Se
$4p$ orbitals we construct a basis set of Wannier functions
using the projection procedure on a local atomic-centered
symmetry-constrained basis set within a window spanning both
the Fe $3d$ and Se $4p$ bands~\cite{WannierH}. To solve the
realistic many-body problem, we employ the continuous-time
hybridization-expansion (segment) quantum Monte Carlo
algorithm~\cite{ctqmc}. In accordance with previous studies
of the pnictides and chalcogenides, we use the average
Coulomb interaction $U = 3.5$~eV and Hund’s exchange
$J = 0.85$~eV parameters for the Fe $3d$ orbitals~\cite{U_in_superconductors, FeSe_ours}. 
The $U$ and $J$ values are assumed to remain constant upon 
variation of the lattice volume. The Coulomb interaction is 
treated in the density-density approximation. Spin-orbit 
coupling is neglected in these calculations. Furthermore, 
we employ the fully localized double-counting correction, 
evaluated from the self-consistently determined local 
occupations, to account for the electronic interactions 
already described by DFT. To compute spectral properties 
and renormalizations of the effective electron mass of the 
Fe $3d$ orbitals, we calculate the real-axis self-energy 
$\Sigma(\omega)$ using Pad\'e analytical continuation 
procedure~\cite{Pade}. The evolution of magnetic correlations 
under pressure is analyzed by calculating the momentum-resolved 
static susceptibility in the particle-hole bubble approximation:
\begin{equation}
\chi({\bf q})=
-k_{\mathrm{B}}T\sum_{{\bf k},i\omega_{n}}\mathrm{Tr}
\hat G({\bf k},i\omega_{n})
\hat G({\bf k}+{\bf q},i\omega_{n}),
\end{equation}
where $T$ is the temperature, $\omega_{n}$ is the fermionic
Matsubara frequency and $\hat G({\bf k},i\omega_{n})$
is the interacting lattice Green's function computed within
DFT+DMFT.


\section{RESULTS}

\subsection{Crystal structure optimization}

\begin{figure}[t]
\centering
\includegraphics[width=0.48\textwidth,clip=true,angle=-90]{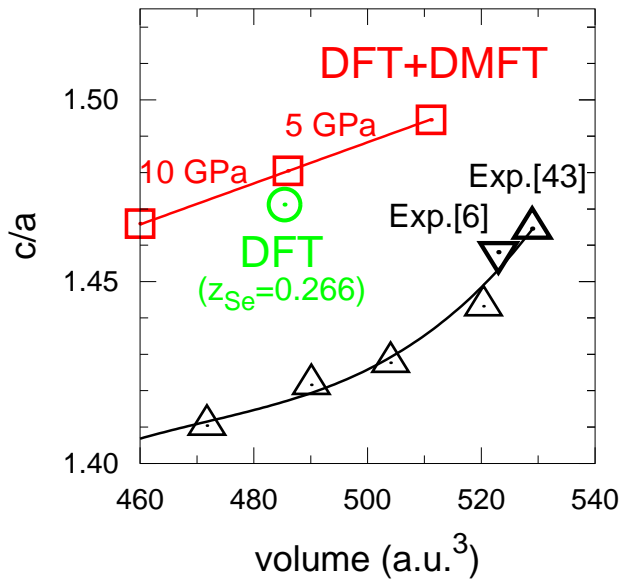}
\caption{(Color online) Evolution of the lattice volume and
$c/a$ ratio of paramagnetic tetragonal FeSe under pressure
obtained by DFT+DMFT at $T = 290$~K (squares) as compared
to the experimental results (triangles~\cite{FeSe_structure_pressure},
inverted triangle~\cite{FeSe_structure}). Bold triangles
mark the cell parameters at ambient pressure. The nm-GGA result
with $z_\mathrm{Se}$ fixed to its experimental value is shown
by a circle. Note that a fully relaxed result obtained within 
nm-GGA would be out of the range of the plot (see text).
}

\label{Fig_1}
\end{figure}

We start by performing a structural optimization of
tetragonal $(P4/nmm)$ FeSe at ambient pressure. To this
end, we compute the total energy as a function of volume,
lattice parameter $c/a$, and the internal position of Se
$z_\mathrm{Se}$ (the space-group symmetry remains unchanged,
$P4/nmm$) using the charge self-consistent DFT+DMFT method.
The equilibrium structural parameters are determined by
calculating the minimum of the total energy functional. 
Our results are summarized in Fig.~\ref{Fig_1}. We note 
that the equilibrium lattice parameters of FeSe computed 
within the nonmagnetic GGA (nm-GGA) using the plane-wave 
pseudopotential approach, i.e., $V = 563$~a.u.$^3$, 
$c/a = 1.671$ and $z_\mathrm{Se} = 0.227$, are substantially 
(by $\sim$$6-15$~\%) off the experimental values; this is 
out of the range of the plot of the $c/a$ ratio vs. volume 
shown in Fig.~\ref{Fig_1}~\cite{FeSe_structure,FeSe_structure_pressure}.
Fixing the $z$ position of Se to its experimental value 
$z_\mathrm{Se} = 0.266$, the equilibrium lattice volume and 
the $c/a$ ratio values calculated within nm-GGA are 
$485$~a.u.$^3$ and $1.471$, respectively, which is about 
$8$~\% lower (volume) and $\sim$$1$~\% larger ($c/a$) than 
those found in experiment (see Fig.~\ref{Fig_1}).

By contrast, our results for the equilibrium lattice parameters
obtained by DFT+DMFT agree well with experiment. In particular,
we find the equilibrium volume as $511$~a.u.$^3$ and the $c/a$
ratio as $1.494$, which agree within $3$~\% with the experimental values.
The calculated Se $z$ fractional position $z_\mathrm{Se} = 0.261$
is in excellent agreement with experiment~\cite{FeSe_structure}. 
Furthermore, while in nm-GGA the results for the lattice parameters 
obtained with and without optimization of $z_{\mathrm{Se}}$ strongly differ, 
this is not the case within DFT+DMFT.
Our result for the (instantaneous) local magnetic moment
$\sqrt{\langle\hat{m}_z^2\rangle}\sim 1.86$~$\mu_{\mathrm{B}}$
is close to that found in the previous charge self-consistent
DFT+DMFT calculations~\cite{FeSe_ours} with $c/a$ fixed to its
experimental value. The calculated bulk modulus is $K_0$$\sim 85$~GPa,
which is comparable with that for iron pnictides. We note that
this is much lower than the result obtained without electronic
correlations ($116$~GPa).

Moreover, we observe that the $c/a$ ratio obtained within
DFT+DMFT is slightly overestimated by about 3~\%. This is
presumably related to the van der Waals (attractive)
interaction between the electrically neutral FeSe layers,
which is not taken into account in our calculations. It has 
been shown previously that in the binary chalcogenides the 
local (repulsive) Coulomb interaction is important for
calculation of the in-plane lattice constant and the volume
of the unit cell~\cite{FeSe_ours}. On the other hand, the
van der Waals attraction was found to correct the overestimation 
of the out-of-plane lattice parameter in DFT~\cite{vdW}.

Next, we calculate the evolution of the lattice parameters
of FeSe under pressure (see Fig.~\ref{Fig_1}). For simplicity,
in these calculations the internal coordinate $z_{\mathrm{Se}}$
was fixed at its experimental value which is very close to
that obtained within DFT+DMFT at ambient pressure. Furthermore,
for subsequent verification we perform an optimization of
$z_{\mathrm{Se}}$ at pressure $\sim$$10$~GPa, which results
in a weak $\sim$$2$~\% increase of $z_{\mathrm{Se}}$, in
agreement with experiment~\cite{FeSe_structure_pressure}.
In Fig.~\ref{Fig_1} we display our results for the evolution
of the $c/a$ ratio and the lattice volume of FeSe upon
compression. Our calculations exhibit a monotonous decrease
of the $c/a$ ratio and the lattice volume of FeSe, which can
be approximated (up to accuracy of the present DFT+DMFT
calculations) by a straight line. The slope of this line, i.e.,
the relative dependence of the $c/a$ on the lattice volume,
is in good agreement with experimental data for the unit cell
parameters at $2-11$~GPa~\cite{FeSe_structure_pressure}.
However, the experimental data show a substantial change 
(kink) in the slope of the $c/a$ ratio vs. lattice volume
at about $2$~GPa ($V \sim 520$ a.u.$^3$), which is not reproduced
in our DFT+DMFT calculations. We attribute this discrepancy
to the van der Waals attraction between the FeSe layers, 
which is not included in the present calculations.

Our results for the lattice parameters of FeSe under pressure
are in overall good agreement with recent experimental data~\cite{FeSe_structure_pressure}.
Moreover, our calculations demonstrate the crucial importance
of electronic correlations in determining the
structural properties of FeSe under pressure. For a pressure of
$\sim$$10$~GPa, we obtain a lattice volume $V \sim 460$~a.u.$^3$
and a ratio $c/a = 1.464$, which is $< 1$~\% larger than the lattice
volume and about $4$~\% larger than $c/a$ observed in
experiment (see Table~\ref{Table1}), respectively. Our result for the local
magnetic moment $1.82$~$\mu_B$ is close to the result obtained
at ambient pressure.

\begin{table}
\caption{
Calculated structural parameters of tetragonal FeSe at ambient
pressure and those at about $\sim$$10$~GPa as obtained by nm-GGA
and DFT+DMFT in comparison with experiment \cite{FeSe_structure,FeSe_structure_pressure}.
Here $h_{\mathrm{Se}}$ is the height of Se above the basal Fe
plane, $d_{\mathrm{Fe-Se}}$ and $d_{\mathrm{Fe-Fe}}$ are the
distances between the corresponding ions. a.p. - results obtained at ambient pressure.}
\begin{tabular}{c|cccccc}
\hline
\hline
$p$        & $V$        & $c/a$     & $z_{\mathrm{Se}}$ & $d_{\mathrm{Fe-Se}}$ & $d_{\mathrm{Fe-Fe}}$ & $h_{\mathrm{Se}}$ \\
(GPa)      & (a.u.$^3$) &           &                   & (a.u.)               & (a.u.)               & (a.u.)            \\
\hline
           &         \multicolumn{6}{c}{DFT+DMFT}                                         \\
a.p.       &  $511$     & $1.494$   & $0.261$           & $4.438$   & $4.950$  & $2.729$  \\
$10$       &  $460$     & $1.464$   & $0.280$           & $4.395$   & $4.807$  & $2.787$  \\
\hline
           &         \multicolumn{6}{c}{nm-GGA}                                           \\
a.p.       &  $563$     & $1.671$   & $0.227$           & $4.366$   & $4.920$  & $2.638$  \\
$10$       &  $433$     & $1.402$   & $0.282$           & $4.307$   & $4.781$  & $2.670$  \\
\hline
           &         \multicolumn{6}{c}{Exp. (room temperature)}                     \\
a.p. \cite{FeSe_structure}        &  $523$     & $1.458$   & $0.266$           & $4.498$   & $5.024$  & $2.759$  \\
a.p. \cite{FeSe_structure_pressure}  &  $529$     & $1.464$   & $0.288$           & $4.657$   & $5.036$  & $3.001$  \\
$8.5$ \cite{FeSe_structure_pressure}     &  $456$     & $1.406$   & $0.319$           & $4.614$   & $4.858$  & $3.080$  \\
$11$ \cite{FeSe_structure_pressure}      &  $447$     & $1.397$   & $0.306$           & $4.487$   & $4.824$  & $2.915$  \\
\hline
\hline
\end{tabular}
\label{Table1}
\end{table}


\subsection{Spectral properties and Fermi surface under pressure}

\begin{figure}[t]
\centering
\includegraphics[width=0.47\textwidth,clip=true,angle=0]{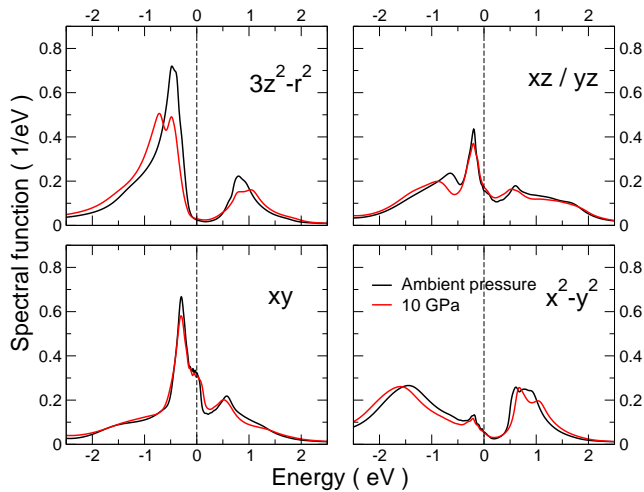}
\caption{(Color online) Orbitally-resolved Fe $3d$ spectral
functions of FeSe obtained within DFT+DMFT at $T = 290$~K for
ambient pressure and for about $10$~GPa. The Fermi
energy ($E_{\mathrm F} = 0$~eV) is shown by a vertical line.
}
\label{Fig_2}
\end{figure}

We now compute the evolution of the spectral properties of
FeSe under pressure. Our results for the Fe $3d$ spectral
functions obtained by DFT+DMFT at ambient pressure and upon
compression to $\sim$$10$ GPa are shown in Fig.~\ref{Fig_2}.
The former results are in good quantitative agreement with
our previous DFT+DMFT study of FeSe~\cite{FeSe_ours}. In
particular, all five Fe $3d$ orbitals form a band with a
predominant contribution at the Fermi level originating from
the $t_2$ states. Moreover, in accordance with previous
investigations, we observe a reduction of the Fe $3d$ bandwidth
(in comparison to the DFT results) near the Fermi energy
caused by electronic correlations. The lower Hubbard band
associated with the Fe $d_{xy}$ and $d_{xz/yz}$ orbitals is
located at about $-1.5$~eV~\cite{FeSe_ours,FeSe_LHB_2,FeSe_arpes_2}. 
The Fe $d_{xy}$ and $d_{xz/yz}$ spectral functions exhibit a 
well-defined quasiparticle peak located below the Fermi level 
at about $0.2$~eV, originating from the van Hove singularity 
of the $d_{xy}$ and $d_{xz/yz}$ bands at the M point. Upon 
compression, we find no substantial spectral weight transfer, 
in remarkable contrast to the DFT+DMFT results obtained upon 
\emph{expansion} of the lattice. Our results exhibit a weak 
increase of the bandwidth and a subtle splitting of the peaks 
associated with the $e$ states located at $\sim$$1$~eV above 
and below the Fermi level. However, the overall shape of the 
spectral functions shows only a weak pressure dependence. 
Under pressure, our results reveal no substantial reduction 
of the Fe $3d$ spectral weight at the Fermi level, which is
less than 1\%. These results agree well with an analysis of 
the band mass enhancement
${m^*}/{m}=1-\partial \mathrm{ Re }\Sigma(\omega)/\partial\omega|_{\omega=0}$,
which provides a quantitative measure of the correlation strength.
We calculate the derivative
$\partial \mathrm{Re}\Sigma(\omega)/\partial\omega|_{\omega=0}=\partial \mathrm{Im}\Sigma(i\omega)/\partial i\omega|_{i\omega=0}$
employing Pad\'e extrapolation of the self-energy $\Sigma(\omega)$
to $\omega=0$~eV. In particular, under pressure of $\sim$$10$~GPa 
we obtain a $\sim$$5$~\% reduction of $m^*/m$ for the $e$ 
states and less than $1$~\% for the $t_2$ states, as compared 
to the ambient pressure.

\begin{figure}[t]
\centering
\includegraphics[width=0.45\textwidth,clip=true,angle=-90]{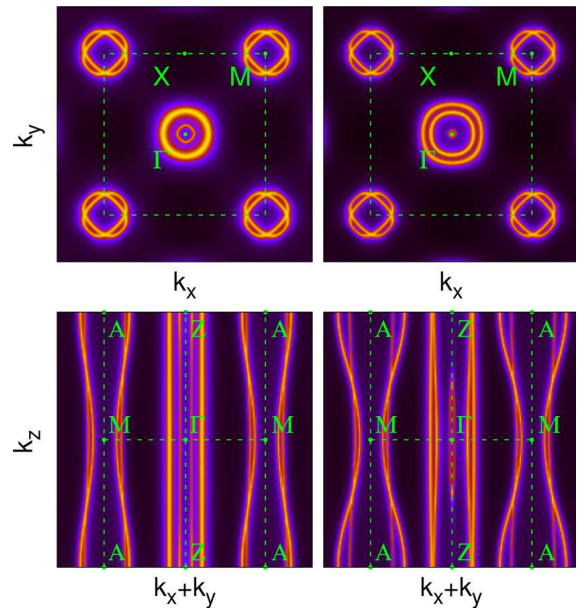}
\caption{(Color online) Fermi surface sections in the $k_z=0$
(top) and $k_x=k_y$ (bottom) planes of the tetragonal Brillouin
zone of FeSe at ambient pressure (left column) and at $p = 10$~GPa
(right column) as computed by DFT+DMFT at $T = 290$~K. Under 
pressure the inner hole-like Fermi surface centered at the
$\Gamma$ point collapses halfway between the $\Gamma$ and Z
points, resulting in a closed elliptic-like three-dimensional 
FS (bottom-right panel).}
\label{Fig_3}
\end{figure}

To proceed further, we calculate the {\bf k}-resolved
spectral functions of FeSe. In Fig.~\ref{Fig_3} we present
our results for the Fermi surface (FS) of FeSe calculated
for $k_z=0$ and $k_x=k_y$ by integration of the spectral
weight
$A({\bf k},\omega) \equiv -\frac{1}{\pi}\mathrm{Im}[G({\bf k},\omega)]$
over a $\pm 5$~meV energy window around the Fermi level.
In agreement with previous studies, at ambient pressure we
find two intersecting elliptical electron Fermi surfaces
centered at the Brillouin zone M point, associated with
the $d_{xy}$ and $d_{xz/yz}$ orbitals. In addition, there
are three concentric hole-like pockets at the $\Gamma$
point (the two outer hole pockets are degenerate). 
The FS is seen to be remarkably expanded as compared to 
the prediction of the low-temperature ARPES 
measurements~\cite{review_chalcogen_recent,FeSe_arpes_1,FeSeTe_arpes}. 
It was indeed proposed that this is related to a ``blue-red shift'' 
problem~\cite{FeSe_arpes_recent}, i.e., unappropriate 
shifting upward of the electron bands at the M and downward 
of the hole-like bands at the $\Gamma$ point, presumably 
due to a momentum dependence of the real part of self-energy 
$\Sigma({\bf k}, \omega)$ (the latter by construction is 
local, {\bf k}-independent $\Sigma(\omega)$, in our single-site 
DFT+DMFT calculations). We note however that our DFT+DMFT 
calculations are carried out at $T=290$~K, whereas a direct 
comparison of our result for the FS with recent room-temperature 
ARPES measurements~\cite{FeSe_arpes_recent} shows good 
quantitative agreement.

We find that bulk FeSe is a correlated metal with coherent
quasiparticle behavior at the Fermi level (well-defined FS),
which exhibits incoherent spectral weight at higher binding
energies, in agreement with recent experiments~\cite{FeSe_arpes_recent}.
We note that at ambient pressure, a structural optimization
including relaxation of the $z_{\mathrm{Se}}$ coordinate
neither affects the topology of the FS nor the in-plane
nesting vector $(\pi,\pi)$, connecting the electron-like and
hole-like FS pockets. The latter is indicative
of stripe-type spin excitations, consistent with a
$s^{\pm}$ pairing symmetry \cite{FeSe_spm}. We also note that
the size of the FS pockets depends only weakly on the structural
optimization of the lattice.

\begin{figure}[t]
\centering
\includegraphics[width=0.45\textwidth,clip=true,angle=0]{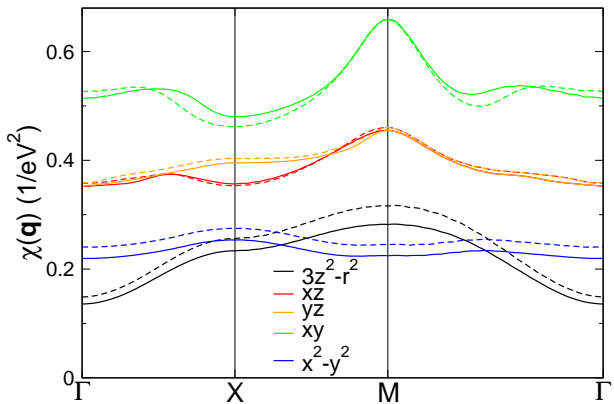}
\caption{(Color online) Orbital contributions to the static spin
susceptibility $\chi({\bf q})$ of paramagnetic FeSe at ambient
pressure (broken line) and $p = 10$~GPa (solid line) calculated
along the $\Gamma$-X-M-$\Gamma$ path using DFT+DMFT.
}
\label{Fig_4}
\end{figure}

Upon compression of the FeSe lattice, we observe a remarkable
reconstruction of the electronic structure and of the FS, i.e., 
a Lifshitz transition (see Fig.~\ref{Fig_3}, right
panel). In particular, the inner hole-like FS centered at
the $\Gamma$ point collapses halfway between the
$\Gamma$ and Z points, resulting in a closed elliptic-like
three-dimensional FS, elongated in the $k_z$ direction,
around $\Gamma$. At the same time, the topology of the
electron-like FS at the M point shows no qualitative change,
while its cross-section in the $(k_x,k_y)$ plane now exhibits
a substantial variation for different $k_z$. In fact,
the elongation of the FS ellipses in the $(k_x,k_y)$ cross-section
is small for $k_z=0$ and large for $k_z=\pi$. Overall, our
results clearly show that bulk FeSe under pressure undergoes
a Lifshitz transition which is associated with a two- to 
three-dimensional crossover of the FS.

Bulk FeSe at about 10~GPa is seen to be a moderately
correlated metal with coherent quasiparticle behavior at 
the Fermi level. The latter makes the electronic properties 
of FeSe particularly interesting since, both under positive
and negative (chemical) pressure, it exhibits a Lifshitz
transition. Interestingly, according to experiments $T_c$ 
is boosted up to $\sim$$37$~K in FeSe under pressure of about 
9~GPa~\cite{FeSe_magn_sc_pressure} and up to $\sim$$17$~K in 
FeSe$_{1-x}$Te$_x$ upon negative chemical pressure caused by
substituting Se by Te~\cite{Tc_negative_pressure}. However, 
on a qualitative level, the Lifshitz transition upon positive
pressures does not result in a coherence-incoherence crossover
as it was found upon expansion of the lattice in Refs.\cite{FeSe_ours,FeSeTe_arpes}.

\begin{figure}[t]
\centering
\includegraphics[width=0.45\textwidth,clip=true,angle=-90]{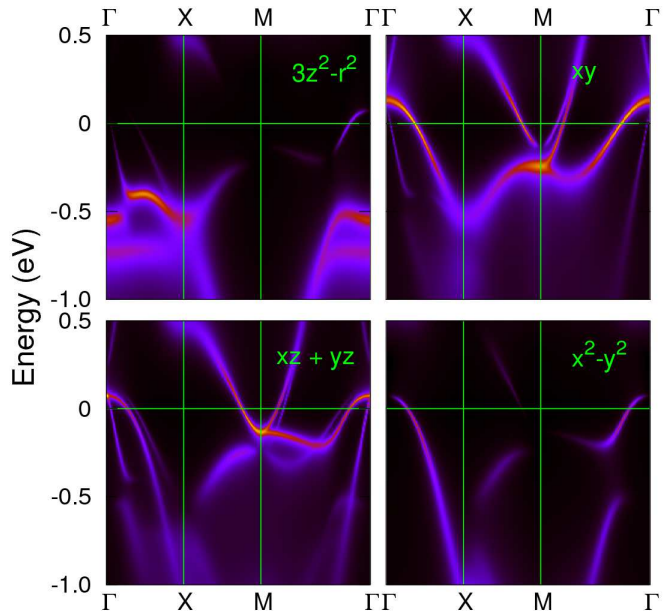}
\caption{(Color online) Orbitally-resolved band structure of
FeSe along the $\Gamma$-X-M-$\Gamma$ path as obtained by DFT+DMFT
at $T = 290$~K and $ p = 10$~GPa.
}
\label{Fig_5}
\end{figure}

The change of the FS results in a corresponding change of
magnetic correlations due to a suppression of the in-plane
nesting vector $Q_m=(\pi,\pi)$. In particular, our results
exhibit a small reduction of the magnetic
correlations under (positive) pressure, i.e., $\chi({\bf q})$
shows a small decrease due to the $e$ states compared to
the results at ambient pressure. At the same time the in-plane nesting
$Q_m$ vector remains of the $(\pi,\pi)$-type which is consistent
with stripe-type spin fluctuations (see Fig.~\ref{Fig_4}).
This is in stark contrast to the $(\pi,\pi)$-to-$(\pi,0)$
reconstruction of the magnetic correlations in FeSe upon \emph{expansion}
of the lattice (negative chemical pressure in Fe(Se,Te))
previously found in \cite{FeSe_ours,FeSeTe_arpes}. Furthermore,
we note that the nm-GGA results reveal no substantial change
in the FS of FeSe upon moderate compression or
expansion of the lattice. This highlights once more the great importance
of electronic correlations, which are found to be responsible for the 
remarkable reconstruction of the electronic structure of
bulk FeSe, associated with a Lifshitz transition, both upon
expansion \emph{and} compression of the lattice volume.

\begin{figure}[t]
\centering
\includegraphics[width=0.45\textwidth,clip=true,angle=-90]{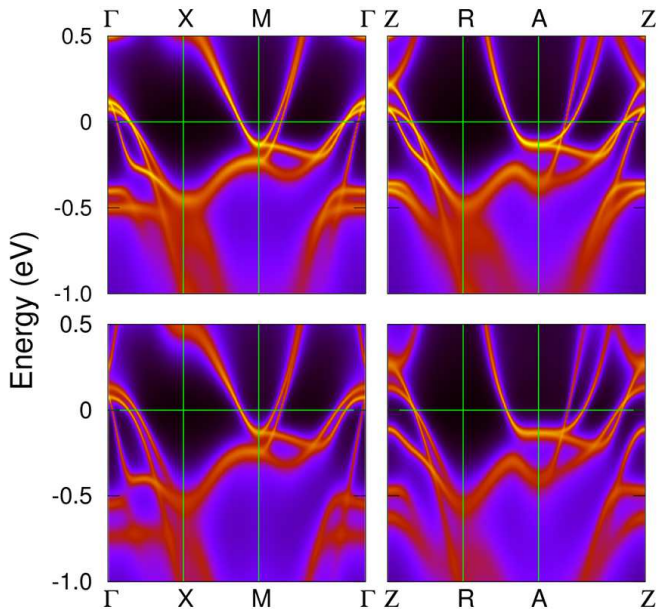}
\caption{(Color online) Band structure of FeSe along the
$\Gamma$-X-M-$\Gamma$ and Z-R-A-Z directions at ambient
pressure (top row) and $p = 10$~GPa (bottom row) as computed
by DFT+DMFT at $T = 290$~K.
}
\label{Fig_6}
\end{figure}

To analyze our results further we calculated the
electronic structure of bulk FeSe at ambient pressure and
at about 10~GPa. In Fig.~\ref{Fig_5} we present our results
for the orbitally-resolved spectral function evaluated in
the $\Gamma$-X-M plane of the Brillouin zone ($k_z=0$).
The total spectral functions of FeSe computed in the
$\Gamma$-X-M ($k_z=0$) and R-Z-A ($k_z=\pi$) planes are
shown in Fig.~\ref{Fig_6}. In agreement with previous
DFT+DMFT calculations, we obtain a substantial renormalization
of the effective crystal-field splitting of the Fe $3d$
bands with respect to the DFT results caused by the strong
energy and orbital dependence of the self-energy~\cite{FeSe_ours,FeSe_Haule_1,FeSe_arpes_2}.
The latter leads to different shifts of the quasiparticle
bands near the Fermi level. In particular, in agreement
with the ARPES measurements, we observe a ``blue-red shift''
of the states near the M point (which are pushed upwards)
and the hole pockets near the $\Gamma$ point (pushed downwards).
Both shift towards the Fermi level (see Fig.~\ref{Fig_6})
due to orbital-selective correlation effects. In addition,
our results clearly distinguish the van Hove singularity at
the M point, associated with the Fe $d_{xy}$ and $d_{xz/yz}$
orbitals (see Fig.~\ref{Fig_5}).

We note that at ambient pressure the spectral function of
FeSe evaluated in the $\Gamma$-X-M ($k_z=0$) and R-Z-A
($k_z=\pi$) planes (see Fig.~\ref{Fig_6}, top row) exhibits
the same structure. It consists of three
hole-like bands near the center (at the $\Gamma$ and Z points)
and two electron-like bands near the Brillouin zone corners
(at the M and A points). By contrast, under pressure of
$\sim$$10$~GPa, we observe an orbital-dependent shift of
the energy bands near the Z point, resulting in a topological
change of the FS (Lifshitz transition), along the $\Gamma$-Z
(R-Z) direction. In particular, the $xy$ band shifts towards
the Fermi level in the $k_z = 0$ plane but still remains
above the Fermi level (at the $\Gamma$ point). On the other
hand, for $k_z=\pi$ the $xy$ band is pushed below the Fermi
level (near the Z point), resulting in the formation of a
closed inner hole pocket in the $\Gamma$-Z direction.
Therefore we conclude that a microscopic mechanism of the
Lifshitz transition which results in a $2$D-$3$D dimensional
crossover in $P4/nmm$ FeSe under pressure is associated with
a pressure-induced shift of the $d_{xy}$-derived band.
We note that the $xy$ band exhibits the largest renormalization
of the quasiparticle mass
$m^{*}/m$~\cite{FeSe_ours,FeSe_LHB_2,FeSe_arpes_2,FeSe_Haule_1}.
Overall, our results show that renormalizations of the band
structure of FeSe under pressure are weaker than those at
ambient pressure as it is expected for a typical correlated
metal.

The reconstruction of the electronic structure and
the evolution of the FS of FeSe under pressure agree reasonably
well with available experiments. Moreover, our results are
in line with the experimentally observed expansion of the FS
of the FeSe$_{1-x}$S$_x$ series upon addition of
S~\cite{FeS_corr_suppress,Watson_prb_15}. In fact, tetragonal
FeS is an isoelectronic counterpart of FeSe with a smaller
lattice volume, due to a smaller ionic radius of S with
respect to Se. Therefore the properties of
FeSe$_{1-x}$S$_x$ can be regarded as those corresponding to FeSe under
positive pressure. With this in mind, our results imply
that FeSe under pressure and, hence, FeSe$_{1-x}$S$_x$ are
less correlated than FeSe, in agreement with recent
studies~\cite{FeS_corr_suppress}. An analysis of
the FSs and magnetic correlations
of FeSe under pressure in terms of $\chi({\bf q})$ suggests that FeSe$_{1-x}$S$_x$
exhibits stripe-type spin excitations with in-plane
$(\pi,\pi)$ nesting. The latter are weakened by the damping
of spin fluctuations upon compression of the lattice~\cite{Watson_prb_15}.

Under pressure, our results reveal the three hole-like
pockets of the FS in the vicinity of the $\Gamma$ point,
with the inner hole-like FS pocket being of a $3$D
character. However, upon further compression it is seen
that the three hole-like FSs collapse to the two FSs,
in agreement with recent studies of FeS~\cite{FeS_Miao}.
Our results suggest that the pressure-induced
increase of the critical temperature in FeSe, which is
boosted to $37$~K by applying a pressure of about
$9$~GPa~\cite{FeSe_magn_sc_pressure}, is not of
spin-fluctuation origin. Indeed, it indicates that the anomalous
behavior of FeSe upon variation of the lattice volume (both 
compression \emph{and} expansion) is associated 
with a crossover between superconductivity mediated by spin-fluctuations
(upon expansion of the lattice in Fe(Se,Te)) and by phonons
(upon compression), respectively. While the low-pressure
superconducting phase of FeSe most likely has an $s^{\pm}$
order parameter, one may expect a change of the symmetry,
e.g., to a $d$-wave order parameter in the high pressure phase \cite{FeSe_spm,Yang_2016}.


\section{CONCLUSION}

By employing a fully charge self-consistent implementation
of the DFT+DMFT method we performed structural optimization
and studied the electronic properties of the tetragonal ($P4/nmm$)
phase of paramagnetic FeSe under pressure at $T=290$~K.
In agreement with previous studies our
results demonstrate that electronic
correlations need to be included to determine the correct lattice parameters of
FeSe both in the equilibrium and upon compression~\cite{FeSe_ours}.
Our results
for the lattice parameters are in overall good agreement with
experimental data. At ambient pressure, the calculated $c/a$
ratio is about $3$ \%  larger than the experimental
value, suggesting a possible importance of van der Waals
attraction between the FeSe layers, which are absent in the
present calculations. Upon compression to $10$~GPa, the $c/a$
ratio and lattice volume show a linear-like decrease by $2$~\%
and $10$~\%, respectively, while the fractional  coordinate
of Se $z_{\mathrm{Se}}$ weakly increases by $\sim$$2$~\%.

Our results for the spectral properties exhibit a rather
weak dependence of the bandwidth on the lattice volume,
associated with a small change of the Fe-Se hybridization
upon a moderate compression of the lattice up to $10$~GPa.
Most importantly, we obtain a drastic reconstruction
of the Fermi surface (FS) topology, i.e., a Lifshitz transition,
indicating a two- to three-dimensional crossover in the FS. 
In particular, the inner hole-like pocket at the
$\Gamma$ point of the FS collapses halfway between the
$\Gamma$ and Z points, showing a three-dimensional character.
We conclude that a microscopic mechanism of the Lifshitz
transition is associated with a pressure-induced shift
of the $d_{xy}$ orbital in the Z point below the Fermi
level. Our results are in good agreement with recent ARPES
measurements for the FeSe$_{1-x}$S$_x$ solid solution,
which is regarded as an isoelectronic and isostructural
(with a smaller lattice volume due to chemical pressure)
counterpart of FeSe. Under pressure, we observe a small
reduction of the quasiparticle mass renormalization $m^*/m$
by about $5$~\% for the $e$ and less than $1$~\% for the
$t_2$ states, as compared to results at ambient pressure.
This implies that FeSe$_{1-x}$S$_x$ is less correlated
than FeSe, in agreement with recent studies~\cite{FeS_corr_suppress,Watson_prb_15,FeS_Miao}.

The behavior of the momentum-resolved magnetic susceptibility
$\chi({\bf q})$ under pressure shows no topological change
of magnetic correlations under pressure. The latter suggests
that both FeSe under pressure and FeSe$_{1-x}$S$_x$ exhibit 
stripe-type spin excitations. Moreover, an analysis of
the FS and magnetic correlations exhibits a small
reduction of the degree of the in-plane $(\pi,\pi)$ nesting,
which implies that magnetic correlations are weakened due to
the damping of spin fluctuations under pressure. We speculate
that the anomalous behavior of FeSe upon variation of the
lattice volume, both under expansion and compression, is associated with a crossover between 
superconductivity mediated by spin-fluctuations (upon expansion of the lattice
in Fe(Se,Te)) and by phonons (upon compression), respectively.


\section{ACKNOWLEDGMENTS}

The work was supported by the Russian Scientific Foundation
(Project No. 14-22-00004). S.L.S., V.I.A., and I.L. are grateful
to the Center for Electronic Correlations and Magnetism,
University of Augsburg, for hospitality.



\begin{thebibliography}{99}


\bibitem{review_chalcogen_recent}
X. Liu, L. Zhao, S. He, J. He, D. Liu, D. Mou, B. Shen,
Y. Hu, J. Huang, and X. J. Zhou,
J. Phys.: Condens. Matter {\bf 27}, 183201 (2015);
%
Y. V. Pustovit and A. A. Kordyuk,
Low Temp. Phys. {\bf 42}, 995 (2016);
%
A. I. Coldea and M. D. Watson,
Annu. Rev. Condens. Matter Phys. {\bf 9}, 125 (2018).
%
A. E. B\"ohmer and A. Kreisel,
J. Phys.: Condens. Matter {\bf 30}, 023001 (2018).


\bibitem{review_fe_supercond_general}
J. Paglione and R. L. Greene,
Nat. Phys. {\bf 6}, 645 (2010);
%
D. N. Basov and A. V. Chubukov,
Nat. Phys. {\bf 7}, 272 (2011);
%
G. R. Stewart,
Rev. Mod. Phys. {\bf 83}, 1589 (2011);
%
P. Dai, J. Hu, and E. Dagotto,
Nat. Phys. {\bf 8}, 709 (2012);
%
Q. Si, R. Yu, and E. Abrahams,
Nat. Rev. Mater. \textbf{1}, 16017 (2016);


\bibitem{pnictide_discovery}
Y. J. Kamihara, T. Watanabe, M. Hirano, and H. Hosono,
J. Am. Chem. Soc. {\bf 130}, 3296 (2008);
%
Z. A. Ren, W. Lu , J. Yang, W. Yi, X. L. Shen, Z. C. Li,
G. C. Che, X. L. Dong, L. L. Sun, F. Zhou, Z. X. Zhao,
Chin.Phys. Lett. {\bf 25}, 2215 (2008);
%
X. H. Chen, T. Wu, G. Wu, R. H. Liu, H. Chen, and D. F. Fang,
Nature (London) {\bf 453}, 761 (2008).


\bibitem{Superconductivity_FeSe}
F. C. Hsu, J. Y. Luo, K. W. Yeh, T. K. Chen, T. W. Huang, P. M. Wu,
Y. C. Lee, Y. L. Huang, Y. Yi. Chu, D. C. Yan and M. K. Wu,
Proc. Natl. Acad. Sci. U.S.A. {\bf 105}, 14262 (2008).


\bibitem{FeSe_magn_sc_pressure}
S. Medvedev, T. M. McQueen, I. A. Troyan, T. Palasyuk, M. I.
Eremets, R. J. Cava, S. Naghavi, F. Casper, V. Ksenofontov, G.
Wortmann, and C. Felser,
Nat. Mater. {\bf 8}, 630 (2009);
Q. Wang, Y. Shen, B. Pan, X. Zhang, K. Ikeuchi, K. Iida, 
A. D. Christianson, H. C. Walker, D. T. Adroja, M. Abdel-Hafiez, 
X. Chen, D. A. Chareev, A. N. Vasiliev, and J. Zhao,
Nat. Commun. {\bf 7}, 12182 (2016).


\bibitem{FeSe_structure}
S. Margadonna , Y. Takabayashi, M. T. McDonald, K. Kasperkiewicz,
Y. Mizuguchi, Y. Takano, A. N. Fitch, E. Suard, K. Prassides,
Chem. Commun. (Cambridge) {\bf 43}, 5607 (2008);
%
M. C. Lehman, A. Llobet, K. Horigane, and D. Louca,
J. Phys. Conf. Ser. {\bf 251}, 012009 (2010).


\bibitem{FeSe_tetr_ort}
T. M. McQueen, A. J. Williams, P. W. Stephens, J. Tao,
Y. Zhu, V. Ksenofontov, F. Casper, C. Felser and R. J. Cava,
Phys. Rev. Lett. {\bf 103}, 057002 (2009).


\bibitem{FeSe_magn_1GPa}
M. Bendele, A. Amato, K. Conder, M. Elender, H. Keller,
H.-H. Klauss, H. Luetkens, E. Pomjakushina, A. Raselli
and R. Khasanov,
Phys. Rev. Lett. {\bf 104}, 087003 (2010);
%
M. Bendele M, A. Ichsanow, Y. Pashkevich, L. Keller,
T. Str\"assle, A. Gusev, E. Pomjakushina, K. Conder,
R. Khasanov and H. Keller,
Phys. Rev. B {\bf 85}, 064517 (2012).


\bibitem{FeSe_sc_enhance_press}
Y. Mizuguchi, F. Tomioka, S. Tsuda, T. Yamaguchi and
Y. Takano,
Appl. Phys. Lett. {\bf 93}, 152505 (2008);
%
S. Margadonna, Y. Takabayashi, Y. Ohishi, Y. Mizuguchi,
Y. Takano, T. Kagayama, T. Nakagawa, M. Takata and
K. Prassides,
Phys. Rev. B {\bf 80} 064506 (2009);
%
G. Garbarino, A. Sow, P. Lejay, A. Sulpice, P. Toulemonde,
M. Mezouar and M. Nunez-Regueiro,
Europhys. Lett. {\bf 86} 27001 (2009);
%
S. Masaki, H. Kotegawa, Y. Hara, H. Tou, K. Murata,
Y. Mizuguchi and Y. Takano,
J. Phys. Soc. Japan {\bf 78} 063704 (2009);
%
H. Okabe, N. Takeshita, K. Horigane, T. Muranaka and
J. Akimitsu,
Phys. Rev. B {\bf 81} 205119 (2010).


\bibitem{Tc_negative_pressure}
M. H. Fang, H. M. Pham, B. Qian, T. J. Liu, E. K. Vehstedt,
Y. Liu, L. Spinu, and Z. Q. Mao,
Phys. Rev. B {\bf 78}, 224503 (2008);
%
B. C. Sales, A. S. Sefat, M. A. McGuire, R. Y. Jin, D. Mandrus,
and Y. Mozharivskyj,
Phys. Rev. B {\bf 79}, 094521 (2009);
%
A. Martinelli, A. Palenzona, M. Tropeano, C. Ferdeghini, M. Putti,
M. R. Cimberle, T. D. Nguyen, M. Affronte, and C. Ritter,
ibid. {\bf 81}, 094115 (2010);
%
V. Tsurkan, J. Deisenhofer, A. G\"unther, Ch. Kant, H.-A. Krug von Nidda,
F. Schrettle, and A. Loidl,
Eur. Phys. J. B {\bf 79}, 289 (2011);
%
U. R. Singh, S. C. White, S. Schmaus, V. Tsurkan, A. Loidl,
J. Deisenhofer, and P. Wahl,
Phys. Rev. B {\bf 88}, 155124 (2013).


\bibitem{FeS_superconductivity}
X. F. Lai, H. Zhang, Y. Q. Wang, X. Wang, X. Zhang,
J. H. Lin, and F. Q. Huang,
J. Am. Chem. Soc. {\bf 137}, 10148 (2015);
%
U. Pachmayr, N. Fehn, and D. Johrendt,
Chem. Commun. {\bf 52}, 194 (2016).


\bibitem{FeSe_intercalation}
M. Burrard-Lucas, D. G. Free, S. J. Sedlmaier, J. D. Wright,
S. J. Cassidy, Y. Hara, A. J. Corkett, T. Lancaster, P. J. Baker,
S. J. Blundell, and S. J. Clarke,
Nat. Mater. {\bf 12}, 15 (2013).


\bibitem{FeSe_monolayer}
S. Tan, Y. Zhang, M. Xia, Z. Ye, F. Chen, X. Xie, R. Peng, D.
Xu, Q. Fan, H. Xu, J. Jiang, T. Zhang, X. Lai, T. Xiang, J. Hu,
B. Xie, and D. Feng,
Nat. Mater. {\bf 12}, 634 (2013).


\bibitem{FeSe_pi_pi}
M. D. Lumsden, A. D. Christianson, D. Parshall, M. B. Stone,
S. E. Nagler, G. J. MacDougall, H. A. Mook, K. Lokshin,
T. Egami, D. L. Abernathy, E. A. Goremychkin, R. Osborn,
M. A. McGuire, A. S. Sefat, R. Jin, B. C. Sales, and D. Mandrus,
Phys. Rev. Lett. {\bf 102}, 107005 (2009);
%
Y. Qiu, W. Bao, Y. Zhao, C. Broholm, V. Stanev, Z. Tesanovic,
Y. C. Gasparovic, S. Chang, Jin Hu, Bin Qian, Minghu Fang,
and Zhiqiang Mao,
Phys. Rev. Lett. {\bf 103}, 067008 (2009);
%
M. D. Lumsden, A. D. Christianson, E. A. Goremychkin,
S. E. Nagler, H. A. Mook, M. B. Stone, D. L. Abernathy,
T. Guidi, G. J. MacDougall, C. de la Cruz, A. S. Sefat,
M. A. McGuire, B. C. Sales, and D. Mandrus,
Nat. Phys. {\bf 6}, 182 (2010).


\bibitem{FeTe_pi_0}
W. Bao, Y. Qiu, Q. Huang, M. A. Green, P. Zajdel,
M. R. Fitzsimmons, M. Zhernenkov, S. Chang, M. Fang,
B. Qian, E. K. Vehstedt, J. Yang, H. M. Pham,
L. Spinu, and Z. Q. Mao,
Phys. Rev. Lett. {\bf 102}, 247001 (2009);
%
T. J. Liu, J. Hu, B. Qian, D. Fobes, Z. Q. Mao,
W. Bao, M. Reehuis, S. A. J. Kimber, K. Proke\v s,
S. Matas, D. N. Argyriou, A. Hiess, A. Rotaru,
H. Pham, L. Spinu, Y. Qiu, V. Thampy, A. T. Savici,
J. A. Rodriguez, and C. Broholm,
Nat. Mater. {\bf 9}, 718 (2010);
%
O. J. Lipscombe, G. F. Chen, C. Fang, T. G. Perring,
D. L. Abernathy, A. D. Christianson, T. Egami, N. Wang,
J. Hu, and P. Dai,
Phys. Rev. Lett. {\bf 106}, 057004 (2011).


\bibitem{FeSe_dft}
A. Subedi, L. Zhang, D. J. Singh, and M. H. Du,
Phys. Rev. B {\bf 78}, 134514 (2008);
K.-W. Lee, V. Pardo, and W. E. Pickett,
Phys. Rev. B {\bf 78}, 174502 (2008);
L. Zhang, D. J. Singh, and M. H. Du,
Phys. Rev. B {\bf 79}, 012506 (2009);
D. Guterding, H. O. Jeschke, and R. Valent\'i,
Phys. Rev. B {\bf 96}, 125107 (2017).


\bibitem{FeSe_spm}
I. I. Mazin, D. J. Singh, M. D. Johannes, and M. H. Du,
Phys. Rev. Lett. {\bf 101}, 057003 (2008); A. V. Chubukov, D. V.
Efremov, and I. Eremin, Phys. Rev. B {\bf 78}, 134512 (2008).


\bibitem{FeSe_arpes_1}
A. Tamai, A. Y. Ganin, E. Rozbicki, J. Bacsa, W. Meevasana,
P. D. C. King, M. Caffio, R. Schaub, S. Margadonna,
K. Prassides, M. J. Rosseinsky, and F. Baumberger,
Phys. Rev. Lett. {\bf 104}, 097002 (2010);
J. Maletz, V. B. Zabolotnyy, D. V. Evtushinsky, S. Thirupathaiah,
A. U. B. Wolter, L. Harnagea, A. N. Yaresko, A. N. Vasiliev,
D. A. Chareev, A. E. B\"ohmer, F. Hardy, T. Wolf, C. Meingast,
E. D. L. Rienks, B. B\"uchner, and S. V. Borisenko,
Phys. Rev. B {\bf 89}, 220506 (2014).


\bibitem{FeSeTe_arpes}
Z. K. Liu, R.-H. He, D. H. Lu, M. Yi, Y. L. Chen, M. Hashimoto,
R. G. Moore, S.-K. Mo, E. A. Nowadnick, J. Hu, T. J. Liu, Z. Q.
Mao, T. P. Devereaux, Z. Hussain, and Z.-X. Shen,
Phys. Rev. Lett. {\bf 110}, 037003 (2013);
Z. K. Liu, M. Yi, Y. Zhang, J. Hu, R. Yu, J.-X. Zhu, R.-H. He,
Y. L. Chen, M. Hashimoto, R. G. Moore, S.-K. Mo, Z. Hussain,
Q. Si, Z. Q. Mao, D. H. Lu, and Z.-X. Shen,
Phys. Rev. B {\bf 92}, 235138 (2015).


\bibitem{FeSe_renorm_1}
M. Yi, Z. Liu, Y. Zhang, R. Yu, J. Zhu, J. Lee, R. Moore,
F. Schmitt, W. Li, S. Riggs, J.-H. Chu, B. Lv, J. Hu, M.
Hashimoto, S.-K. Mo, Z. Hussain, Z. Mao, C.-W. Chu, I. Fisher,
Q. Si, Z.-X. Shen, and D. Lu,
Nat. Commun. {\bf 6}, 7777 (2015);
%
L. Fanfarillo, J. Mansart, P. Toulemonde, H. Cercellier, P. L.
F\`evre, F. Bertran, B. Valenzuela, L. Benfatto, and V. Brouet,
Phys. Rev. B {\bf 94}, 155138 (2016).


\bibitem{FeSe_LHB_1}
D. V. Evtushinsky, M. Aichhorn, Y. Sassa, Z.-H. Liu, J. Maletz,
T. Wolf, A. N. Yaresko, S. Biermann, S. V. Borisenko, and
B. B\"uchner, arXiv:1612.02313.


\bibitem{FeSe_arpes_2}
M. D. Watson, S. Backes, A. A. Haghighirad, M. Hoesch,
T. K. Kim, A. I. Coldea, and R. Valent\`i,
Phys. Rev. B {\bf 95}, 081106(R) (2017).


\bibitem{FeSe_ours}
I. Leonov, S. L. Skornyakov, V. I. Anisimov, and D. Vollhardt,
Phys. Rev. Lett. {\bf 115}, 106402 (2015);
S. L. Skornyakov, V. I. Anisimov, D. Vollhardt and I. Leonov,
Phys. Rev. B {\bf 96}, 035137 (2017).


\bibitem{FeSe_LHB_2}
M. Aichhorn, S. Biermann, T. Miyake, A. Georges, and M. Imada,
Phys. Rev. B {\bf 82}, 064504 (2010).


\bibitem{FeSe_V_underestimation}
A. P. Koufos, D. A. Papaconstantopoulos, and M. J. Mehl,
Phys. Rev. B {\bf 89}, 035150 (2014).


\bibitem{dmft}
W. Metzner and D. Vollhardt,
Phys. Rev. Lett. {\bf 62}, 324 (1989);
%
G. Kotliar and D. Vollhardt,
Phys. Today {\bf 57}, 53 (2004);
%
A. Georges, G. Kotliar, W. Krauth, and M. J. Rozenberg,
Rev. Mod. Phys. {\bf 68}, 13 (1996).


\bibitem{dftdmft_nsc}
V. I. Anisimov, A. I. Poteryaev, M. A. Korotin, A. O. Anokhin, and G. Kotliar,
J. Phys. Condens. Matter {\bf 9}, 7359 (1997);
%
G. Kotliar, S. Y. Savrasov, K. Haule, V. S. Oudovenko, O. Parcollet, and C. A. Marianetti,
Rev. Mod. Phys. {\bf 78}, 865 (2006);
%
J. Kune\v s, I. Leonov, P. Augustinsk\' y, V. K\v r\' apek, M. Kollar, and D. Vollhardt,
Eur. Phys. J. Spec. Top. {\bf 226}, 2641 (2017).


\bibitem{FeSe_Haule_1}
S. Mandal, R. E. Cohen, and K. Haule,
Phys. Rev. B {\bf 89}, 220502(R) (2014);


\bibitem{FeS_dft_dmft}
C. Tresca, G. Giovannetti, M. Capone, and G. Profeta,
Phys. Rev. B {\bf 95}, 205117 (2017).


\bibitem{FeSeTe_correlations}
Z. K. Liu, R.-H. He, D. H. Lu, M. Yi, Y. L. Chen, M. Hashimoto,
R. G. Moore, S.-K. Mo, E. A. Nowadnick, J. Hu, T. J. Liu, Z. Q.
Mao, T. P. Devereaux, Z. Hussain, and Z.-X. Shen,
Phys. Rev. Lett. {\bf 110}, 037003 (2013);
Z. K. Liu, M. Yi, Y. Zhang, J. Hu, R. Yu, J.-X. Zhu, R.-H. He,
Y. L. Chen, M. Hashimoto, R. G. Moore, S.-K. Mo, Z. Hussain,
Q. Si, Z. Q. Mao, D. H. Lu, and Z.-X. Shen,
Phys. Rev. B {\bf 92}, 235138 (2015).


\bibitem{FeS_corr_suppress}
P. Reiss, M. D. Watson, T. K. Kim, A. A. Haghighirad,
D. N. Woodruff, M. Bruma, S. J. Clarke, and A. I. Coldea,
Phys. Rev. B {\bf 96}, 121103(R) (2017).


\bibitem{FeS_dft_phonon}
A. Baum, A. Milosavljevi\' c, N. Lazarevi\' c,
M. M. Radonji\' c, B. Nikoli\' c, M. Mitschek,
Z. Inanloo Maranloo, M. \v S\' cepanovi\' c,
M. Gruji\' c-Broj\v cin, N. Stojilovi\' c,
M. Opel, A. Wang, C. Petrovi\' c, Z. V. Popovi\' c,
and R. Hackl,
arXiv:1712.02724.


\bibitem{Watson_prb_15}
M. D. Watson, T. K. Kim, A. A. Haghighirad, S. F. Blake,
N. R. Davies, M. Hoesch, T. Wolf, and A. I. Coldea,
Phys. Rev. B {\bf 92}, 121108(R) (2015).


\bibitem{dftdmft_sc}
I. Leonov, V. I. Anisimov, and D. Vollhardt,
Phys. Rev. B {\bf 91}, 195115 (2015);
%
I. Leonov,
Phys. Rev. B \textbf{92}, 085142 (2015);
%
I. Leonov, L. Pourovskii, A. Georges, and I. A. Abrikosov,
Phys. Rev. B \textbf{94}, 155135 (2016).


\bibitem{FeSe_arpes_recent}
Y. S. Kushnirenko, A. A. Kordyuk, A. V. Fedorov, E. Haubold,
T. Wolf, B. B\" uchner, and S. V. Borisenko,
Phys. Rev. B {\bf 96}, 100504(R) (2017).


\bibitem{Birch}
F. Birch, Phys. Rev. {\bf 71}, 809 (1947).


\bibitem{GGA}
S. Baroni, S. de Gironcoli, A. Dal Corso, and P. Giannozzi,
Rev. Mod. Phys. {\bf 73}, 515 (2001);
P. Giannozzi, S. Baroni, N. Bonini, M. Calandra, R. Car {\it et al.},
J. Phys.:Condens. Matter {\bf 21}, 395502 (2009).


\bibitem{WannierH}
V. I. Anisimov, D. E. Kondakov, A. V. Kozhevnikov, I. A. Nekrasov,
Z. V. Pchelkina, J. W. Allen, S.-K. Mo, H.-D. Kim, P. Metcalf, S. Suga,
A. Sekiyama, G. Keller, I. Leonov, X. Ren, and D. Vollhardt,
Phys. Rev. B {\bf 71}, 125119 (2005);
%
Dm. Korotin, A. V. Kozhevnikov, S. L. Skornyakov, I. Leonov, N. Binggeli,
V. I. Anisimov, and G. Trimarchi,
Eur. Phys. J. B {\bf 65}, 91-98 (2008);
%
G. Trimarchi, I. Leonov, N. Binggeli, Dm. Korotin, and V. I. Anisimov,
J. Phys.: Condens. Matter {\bf 20}, 135227 (2008).


\bibitem{ctqmc}
P. Werner, A. Comanac, L. de Medici, M. Troyer, and A. J. Millis,
Phys. Rev. Lett. {\bf 97}, 076405 (2006);
E. Gull, A. J. Millis, A. I. Lichtenstein, A. N. Rubtsov,
M. Troyer, and P. Werner,
Rev. Mod. Phys. {\bf 83}, 349 (2011).


\bibitem{U_in_superconductors}
K. Haule, J. H. Shim, and G. Kotliar,
Phys. Rev. Lett. {\bf 100}, 226402 (2008);
%
V. I. Anisimov, D. M. Korotin, M. A. Korotin, A. V. Kozhevnikov,
J. Kune\v s, A. O. Shorikov, S. L. Skornyakov, and S. V. Streltsov,
J. Phys. Condens. Matter {\bf 21}, 075602 (2009);
%
M. Aichhorn, L. Pourovskii, V. Vildosola, M. Ferrero,
O. Parcollet, T. Miyake, A. Georges, and S. Biermann,
Phys. Rev. B {\bf 80}, 085101 (2009);
%
S. L. Skornyakov, A. V. Efremov, N. A. Skorikov, M. A. Korotin,
Yu. A. Izyumov, V. I. Anisimov, A. V. Kozhevnikov, and D. Vollhardt,
Phys. Rev. B {\bf 80}, 092501 (2009);
%
S. L. Skornyakov, A. A. Katanin, and V. I. Anisimov,
Phys. Rev. Lett. {\bf 106}, 047007 (2011);
%
Z. P. Yin, K. Haule, and G. Kotliar,
Nat. Mater. {\bf 10}, 932 (2011);
%
Z. P. Yin, K. Haule, and G. Kotliar,
Nat. Phys. {\bf 7}, 294 (2011);
%
M. Aichhorn, L. Pourovskii, and A. Georges,
Phys. Rev. B {\bf 84}, 054529 (2011);
%
J. M. Tomczak, M. van Schilfgaarde, and G. Kotliar,
Phys. Rev. Lett. {\bf 109}, 237010 (2012);
%
Z. P. Yin, K. Haule, and G. Kotliar,
Phys. Rev. B {\bf 86}, 195141 (2012);
%
P. Werner, M. Casula, T. Miyake, F. Aryasetiawan, A. J. Millis,
and S. Biermann,
Nat. Phys. {\bf 8}, 331 (2012);
%
A. Georges, L. de’ Medici, and J. Mravlje,
Annu. Rev. Condens. Matter Phys. {\bf 4}, 137 (2013);
%
M. Hirayama, T. Miyake, and M. Imada,
Phys. Rev. B {\bf 87}, 195144 (2013);
%
C. Zhang, L. W. Harriger, Z. Yin, W. Lv, M. Wang, G. Tan, Y. Song,
D. L. Abernathy, W. Tian, T. Egami, K. Haule, G. Kotliar, and P. Dai,
Phys. Rev. Lett. {\bf 112}, 217202 (2014);
%
A. van Roekeghem, L. Vaugier, H. Jiang, and S. Biermann,
Phys. Rev. B \textbf{94}, 125147 (2016).


\bibitem{Pade}
H. J. Vidberg and J. W. Serene,
J. Low Temp. Phys. {\bf 29}, 179 (1977).


\bibitem{FeSe_structure_pressure}
R. S. Kumar, Yi Zhang, S. Sinogeikin, Y. Xiao, S. Kumar, P. Chow,
A. L. Cornelius and C. Chen,
J. Phys. Chem. B {\bf 114}, 12597 (2010).


\bibitem{vdW}
F. Ricci and G. Profeta,
Phys. Rev. B {\bf 87}, 184105 (2013).


\bibitem{FeS_Miao}
J. Miao, X. H. Niu, D. F. Xu, Q. Yao, Q. Y. Chen, T. P. Ying,
S. Y. Li, Y. F. Fang, J. C. Zhang, S. Ideta, K. Tanaka, B. P. Xie,
D. L. Feng, and F. Chen,
Phys. Rev. B {\bf 95}, 205127 (2017).


\bibitem{Yang_2016}
Y. Yang, W.-S. Wang, H.-Y. Lu, Y.-Y. Xiang, and Q.-H. Wang,
Phys. Rev. B {\bf 93}, 104514 (2016).


\end{thebibliography}
\end{document}